\documentstyle[prl,aps,psfig]{revtex}
\begin{document}
\draft

\twocolumn[\hsize\textwidth\columnwidth\hsize\csname
@twocolumnfalse\endcsname

\widetext
\title{Direct Transition Between a Singlet Mott Insulator 
and  a Superconductor} 
\author{Massimo Capone}
\address{International School for Advanced Studies (SISSA), 
and Istituto Nazionale per la Fisica della Materia (INFM) \\ 
Unit\`a Trieste-SISSA, Via Beirut 2-4, I-34014 Trieste, Italy}
\author{Michele Fabrizio and Erio Tosatti} 

\address{International School for Advanced Studies (SISSA), 
and Istituto Nazionale per la Fisica della Materia (INFM) \\ 
Unit\`a Trieste-SISSA, Via Beirut 2-4, I-34014 Trieste, Italy, 
and International Centre for Theoretical Physics (ICTP), P.O. Box 
586, I-34014 Trieste, Italy }
\date{\today}

\maketitle

\begin{abstract}

We argue that a normal Fermi liquid and a singlet, spin gapped
Mott insulator cannot be continuously connected, and 
that some intermediate phase must intrude between them. By explicitly 
working out a case study where the singlet insulator is stabilized 
by orbital degeneracy and an inverted Hund's rule coupling, mimicking
a Jahn-Teller effect, we find 
that the intermediate phase is a superconductor. 

\end{abstract}

]

\narrowtext

Understanding metal-insulator transitions driven by electron
correlation, 
the so-called Mott transitions (MT), is a long standing problem.
Renewed attention was recently aroused by the development of the 
so-called Dynamical Mean Field Theory (DMFT)\cite{dmft}, 
a quantum analog of classical mean field theories, 
that treats exactly local temporal fluctuations while freezing
spatial correlations.  
The DMFT has provided us with a complete characterization of the MT 
for the single band Hubbard model (SBHM) within the 
paramagnetic sector, where magnetic instabilities are not allowed.
The insulator-metal MT does not occur by gradual
closing of the Mott gap between the Hubbard bands
in the single-particle spectrum, but it is associated with
the appearance (at a given value $U=U_{c2}$) of 
a narrow ``Kondo'' quasiparticle peak or
resonance at the Fermi level\cite{zrk,kot}.
The peak height  is pinned to the non-interacting value $\rho_0(\mu_0)$,
but its width is finite in the metal and vanishes continuously as the 
MT is approached. For $U < U_{c2}$ 
the metallic state is stable with respect to the  metastable insulating 
solution which exists  for  
$U >  U_{c1}$\cite{zrk,kot}.
Roughly in the interval $U_{c1} < U <U_{c2}$, the 
spectral function presents both the 
broad Hubbard bands and the quasiparticle resonance, thus
combining to some extent the properties of a narrow-band metal
with those of the Mott insulator. For smaller
$U$, the lower and upper Hubbard bands
merge together swamping the resonance away, and the correlated metal
continuously turns into an ordinary metal.
As soon as the constraint to the paramagnetic subspace is released,
antiferromagnetism (AFM) appears, and everything changes. 
Even in partially frustrated models,
where the effects of magnetism are attenuated and
AFM usually sets in only above some $U_M < U_{c2}$,
magnetism still ``contaminates'' the MT. The characteristic
energy scale of the magnetic fluctuations, in proximity
of a continuous MT, necessarily
exceeds the width of the Kondo-resonance, thus affecting
the properties of the ensuing metallic phase, e.g.,turning
it into a spin-density wave.

A new and different situation is to be expected if one could realize 
a singlet, spin-gapped Mott insulator, such as, e.g., a 
spin-liquid insulator.
The entropy of such a state is zero, as opposed to the extensive entropy
of the Mott phase in the SBHM\cite{zrk},
and no symmetry, magnetic or other, is broken.
In this Letter we discuss the consequences of the zero-entropy property on the
MT and on the correlated metal just below the transition point.
To this end, we consider 
a (threefold) orbitally degenerate model arising in the physics of the
alkali-doped fullerides\cite{C60}. For densities
$\langle n\rangle =$2 or 4, the ground state (GS) of this model
system is nonmagnetic
both for small $U$ (metallic), and for large $U$ (singlet
nondegenerate Mott insulator). We show that a 
third, intermediate phase actually intrudes between the
Fermi liquid (FL) metal and the singlet Mott insulator.
Remarkably, the intermediate phase is a superconductor.

The Hamiltonian reads
\begin{equation}
H = \sum_{ij\sigma}\sum_{a,b=1}^3 
t_{ij}^{ab} d^\dagger_{i a\sigma}
d^{\phantom{\dagger}}_{j b \sigma}
+ \sum_i H^{int}_i.
\label{Hamiltonian}
\end{equation}
The interaction term is local
\begin{equation}
H^{int}_i = 
\frac{U}{2} n_i^2 + 
\frac{2J}{3}\left[ \sum_{a=1}^3 n_{ia}^2 
-\sum_{a < b} n_{ia} n_{ib} + 
\frac{3}{4}\sum_{a < b}\Delta_{iab}^2\right], 
\label{Hint}
\end{equation}
where $n_{ia} = \sum_\sigma d_{i a \sigma}^\dagger 
d_{i a \sigma}^{\phantom{\dagger}}$ is the electron number on each orbital
($a=1,2,3$) at site $i$, $n_i=n_{i1}+n_{i2}+n_{i3}$, and
$\Delta_{iab} =\sum_\sigma (
d_{ia\sigma}^\dagger d_{ib \sigma}^{\phantom{\dagger}} + H.c.)$.
Besides the overall on-site repulsion $U$, the model includes a
multiplet-exchange-splitting term $J$.
In presence of a Jahn-Teller coupling to some high frequency on-site
vibration, as shown in Ref. \cite{C60}, $J$ may effectively change sign
from positive (Hund's rule) to negative.
When $\langle n \rangle$
corresponds to 2 or 4 electrons per site (the problem is
electron-hole symmetric around $\langle n\rangle = 3$),
the isolated ion has
a non degenerate GS which is simultaneously a spin and
an orbital singlet. With a negative effective
$J$, it is quite natural that a MT from the paramagnetic metal
towards a non degenerate singlet insulator should take
place increasing $U$\cite{Fabrizio97,C60}.

We carry out a DMFT study of the dynamics of model (\ref{Hamiltonian})
in proximity of the MT for $\langle n\rangle =2,4$ on a 
Bethe lattice with hopping diagonal in the orbital index,
$t_{ij}^{\alpha\beta}= t\delta_{\alpha\beta}$.
The bandwidth is $W=4t$.
The DMFT maps (\ref{Hamiltonian}) onto a threefold-degenerate Anderson
impurity model 
\begin{eqnarray}
H_{I} &=& \sum_{a=1}^3\sum_{n,\sigma}\epsilon_{a n}
c^\dagger_{a n\sigma}c^{\phantom{\dagger}}_{a n\sigma}
+ 
\left(V^{a}_{n}d^\dagger_{a\sigma}
c^{\phantom{\dagger}}_{a n\sigma} + H.c. \right)
\nonumber\\
&& -\mu \sum_{a\sigma} d^\dagger_{a\sigma}d_{a\sigma} + H_{int},
\label{HAIM}
\end{eqnarray}
where $H_{int}$ is the interaction (\ref{Hint}) for the
impurity operators $d_{a\sigma}$'s. The self-consistency
condition is
\begin{equation}
\sum_{n}\frac{\left|V^{a}_{n}\right|^2}
{i\omega - \epsilon_{a n}}
= t^2 G_{a}(i\omega),
\label{self}
\end{equation}
$G_{a}(i\omega)$ being the impurity Green's function for 
orbital $a$\cite{dmft}. We solve the impurity model by exact diagonalization
\cite{caffarel}. The sums in Eqs.(\ref{HAIM}) and (\ref{self})
are truncated to a finite and small number $N_s -1$ of conduction
``bath'' orbitals\cite{footnotens}.
A surprisingly small value
of $N_s$ is indeed enough to capture the qualitative features of the MT
in the SBHM, and $N_s$ =5, the value used here,
gives good quantitative results\cite{caffarel}.
Although the Hamiltonian is O(3) invariant under rotations in
orbital space, we cannot exclude a spontaneous breaking
of this symmetry. 
We explicitly studied solutions with  
orbital symmetry breaking (orbital ordering), but we found that the orbitally
symmetric solution, $G_{a}(i\omega_n) \equiv G(i\omega_n)$,
has always lower energy. In the following, we only consider
such symmetric solutions.

\begin{figure}
\centerline{\psfig{bbllx=80pt,bblly=160pt,bburx=510pt,bbury=640pt,%
figure=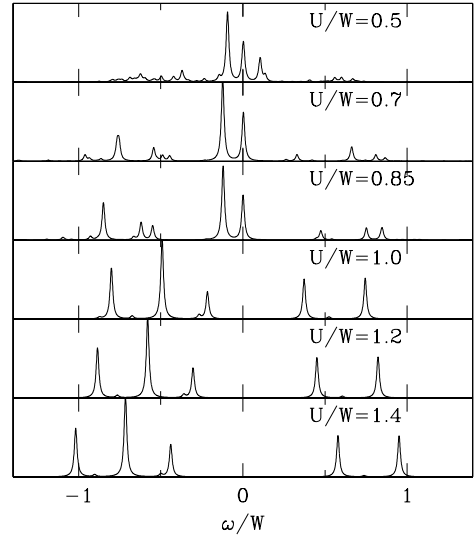,width=68mm,angle=0}}
\caption{$U$ dependence of the single-particle spectral function for
$\langle n\rangle = 4$ and $J/W =-0.02$ in the normal state (with gauge
symmetry enforced). 
\label{green}
}
\end{figure}

\begin{figure}
\centerline{\psfig{bbllx=80pt,bblly=140pt,bburx=510pt,bbury=620pt,%
figure=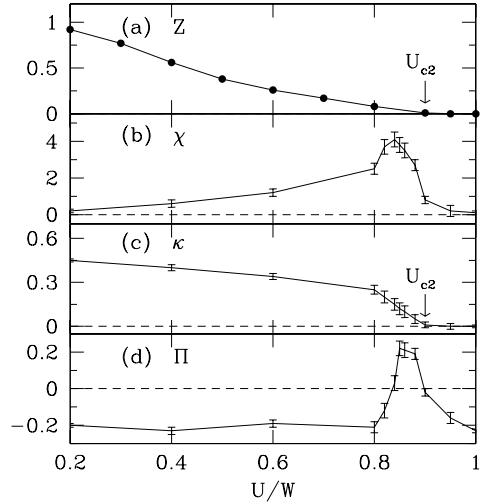,width=60mm,angle=0}}
\caption{$U$ dependence of various relevant quantities in the normal 
state (with gauge symmetry enforced) for $J/U = - 0.02$.
In panel (a) we show the quasiparticle weight $Z$,
in panel (b) the local spin susceptibility $\chi$,
in panel (c) the compressibility $\kappa$,
and in panel (d) the real part of the zero-frequency
pairing susceptibility $\Pi(0)$.
\label{chiekappa}
}
\end{figure}

In order to characterize the metallic phase close to the MT, we compute
the single-particle spectral function
$\rho(\omega) = -1/\pi\ {\cal I}m G(\omega)$,
and  the quasiparticle weight $Z = m/m_{*}$ at the Fermi level, which
is finite in the metallic phase but zero in the insulator, and thus
determines the position of the MT.
Within our numerical accuracy the
vanishing of $Z$  appears to be continuous (see  
Fig. \ref{chiekappa} (a)), as in the SBHM\cite{zrk}.
In Fig. \ref{green} we show the evolution of the spectral
function across the transition, for different values of $U/W$ but at
fixed ratio $J/W= -0.02$.

The metallic phase
close to the transition ($U \lesssim U_{c2}$) shows a coexistence
between high-energy insulating features (Hubbard bands), and the low-energy
metal feature, the Kondo resonance, whose width vanishes
at the transition. In addition, the Hubbard
bands display a well defined multiplet structure, absent in the SBHM,
and typical of the orbitally degenerate site.
Surprisingly, the linewidth of these atomic-like excitations is not set
by the bandwidth as in  the SBHM\cite{zrk},
but by a much smaller energy
scale, as if a kind of motional line narrowing were effectively at work.
The coexistence of atomic and metallic features is therefore much more
striking for our orbitally degenerate model than for the SBHM.
As in  the SBHM,
the single-particle density of states (DOS) at the chemical potential $\mu$
must coincide with the bare DOS $\rho_0(\mu)$. Moreover,
as a consequence of Luttinger's theorem, the  chemical potential
must coincide with the bare one, so that $\rho(\mu) = \rho_0(\mu_0)$
for every $U$, and the MT occurs by
narrowing of the resonance peak. The peak width
$E_F^* \simeq ZW/2$,
may be viewed as an effective Fermi energy for the interacting system,
or a Kondo temperature $T_K$, below which the system may be described as a
FL. As a consequence, the entropy
$S(T)/N\sim T/T_K$ for $T\leq T_K$, so that $S(T_K)$
is of order one up to $U_{c2}$, where
$T_K$ vanishes. By continuity, one can expect this entropy
to be released at the MT. This is indeed what happens in the SBHM,
where the entropy of the paramagnetic
insulator is $S/N = \log{2}$. On the other hand, this is incompatible
with our non-degenerate insulator, which has zero entropy. Therefore,
it is hard to understand how a FL  solution could continuously
connect to the Mott insulator, despite the numerical evidence
strongly pointing
towards a continuous transition. This suggests that the metallic
solution may be metastable close to the MT, and that some
new phase may appear. 

The above heuristic arguments, which apply in any dimension for a
Brinkman-Rice type transition, where the $m_*$ diverges, can be generalized
to other cases where the MT occurs towards a zero-entropy insulator.
The main inadequacy of the metallic solution apparently lies in the constraint
imposed by the Luttinger theorem. Without it, 
the chemical potential could gradually move towards the edge of
the DOS, thus allowing to smoothly connect the
metal with a zero-entropy insulator, as for a metal to a band insulator
transition. This scenario would imply a break-down of the 
Luttinger theorem before the MT.
If  the Mott insulator would break some symmetry of the
Hamiltonian, e.g., the spin SU(2) symmetry,
the metallic phase close to the MT would likely break the same symmetry,
allowing the transition to become of the
metal-band insulator type. This is what happens in the
Hubbard model on a two-dimensional triangular lattice\cite{Massimo},
and also corresponds to what is generally observed experimentally\cite{V2O3}.
If, as in our system, the
insulator does not break any symmetry, it is less clear
what kind of instability to expect.
The presence of orbital degrees of freedom could suggest a 
possible orbital ordering, but, 
as we mentioned before, solutions with broken orbital 
symmetry have always higher energies with respect to 
orbitally symmetric solutions, 
ruling out the possibility of orbital ordering.
Moreover, being  the insulator a spin
singlet, there is no reason to expect a magnetic instability in the metal,
as confirmed by  the vanishing of the local spin susceptibility at the MT.

The only symmetry left, which we have so far enforced in solving the
self consistency equation, is the U(1) gauge symmetry.
A deeper analysis of the metallic phase in the framework of
Landau FL theory suggests that U(1) symmetry might
be spontaneously broken. In our orbitally degenerate model, the
Landau functional must be invariant under spin SU(2) and orbital O(3)
symmetries. We define the local densities
in momentum space
\[
\begin{array}{l}
n_k = \sum_{a=1}^3\sum_{\alpha=\uparrow,\downarrow}
d^\dagger_{a\alpha k}d^{\phantom{\dagger}}_{a\alpha k}, \\
\vec{\sigma}_k = \sum_{a=1}^3\sum_{\alpha,\beta=\uparrow,\downarrow}
d^\dagger_{a\alpha k}\,\vec{\sigma}_{\alpha\beta}\,
d^{\phantom{\dagger}}_{a\beta k}, \\
l_{i,k} = \sum_{a,b=1}^3\sum_{\alpha=\uparrow,\downarrow}
d^\dagger_{a\alpha k}\,L_{i,ab}\,d^{\phantom{\dagger}}_{b\alpha k}, \\
\vec{l}_{i,k} = \sum_{a,b=1}^3\sum_{\alpha,\beta=\uparrow,\downarrow}
d^\dagger_{a\alpha k}\,L_{i,ab}\,\vec{\sigma}_{\alpha\beta}\,
d^{\phantom{\dagger}}_{b\beta k}, \\
\end{array}
\]
where $\vec{\sigma}$ are Pauli matrices, and $L_i$ ($i=x,y,z$) are
the angular momentum operators in the $L=1$ representation
$d_1\sim Y_{1,x}$, $d_2\sim Y_{1,y}$ and $d_3\sim Y_{1,z}$.
We also need to introduce five additional densities
$w_{i,k} = \sum d^\dagger_{a\alpha k}W_{i,ab}
d^{\phantom{\dagger}}_{b\alpha k}$, $i=1,\dots,5$, where the
matrices $W_i$'s are $\sqrt{3}(L_x^2-L_y^2)$,
$\sqrt{3}(L_xL_y+L_yL_x)$,
$\sqrt{3}(L_yL_z+L_zL_y)$,
$\sqrt{3}(L_zL_x+L_xL_z)$ and $(L_x^2+L_y^2-2L_z^2)$, respectively, as
well as their spin antisymmetric counterparts
$\vec{w}_{i,k} = \sum
d^\dagger_{a\alpha k}W_{i,ab}\vec{\sigma}_{\alpha\beta}
d^{\phantom{\dagger}}_{b\beta k}$.
In terms of these densities, the Landau functional reads
\begin{eqnarray}
E &=& \frac{1}{2}\sum_{k,k'} f^S_{k k'}\, n_k n_{k'}
+ f^A_{kk'}\, \vec{\sigma}_{k}\cdot \vec{\sigma}_{k'}
\nonumber\\
&+&\frac{1}{2}\sum_{i=x,y,z}\sum_{k,k'} h^S_{k k'}\, l_{i,k} l_{i,k'}
+ h^A_{kk'}\, \vec{l}_{i,k}\cdot \vec{l}_{i,k'}
\nonumber\\
&+&\frac{1}{2}\sum_{i=1}^5\sum_{k,k'} g^S_{k k'}\, w_{i,k} w_{i,k'}
+ g^A_{kk'}\, \vec{w}_{i,k}\cdot \vec{w}_{i,k'}.
\label{ELandau}
\end{eqnarray}
The local spin susceptibility is given by $\chi/\chi_0 = m_*/[m(1+F_0^A)]$,
where $\chi_0$ is the bare susceptibility, $F_0^A=6 V \rho_0 f^A_0 m_*/m$,
$f^A_0$ being the $l=0$ Legendre transform of the Landau parameter.
Since $\chi$ actually vanishes at the transition, as shown in
Fig.  \ref{chiekappa} (b), then the
antisymmetric Landau parameter $F_0^A$ is positive and
diverges at the MT, probably as $m_*^2$. This is surprising since,
for small $U\gg |J|$, $f_0^A = -U/3 - 10/9 J$ is negative. 
All local orbital and spin orbital susceptibilities
vanish at the transition, since the insulator has a total gap for spin and
orbital excitations, so that the Landau parameters
$H_0^S$, $H_0^A$, $G_0^S$ and $G_0^A$ all diverge from the positive side.
In addition, similarly to the SBHM, the compressibility
$\kappa/\kappa_0 = m_*/[m(1+F_0^S)]$, vanishes like $m_{*}^{-1}$
at the transition (Fig. \ref{chiekappa} (c)). Hence
$F_0^S\sim m_*^2$ also diverges. As a result, the (intra-orbital) 
quasiparticle scattering amplitude in the singlet-channel
\[
A_s \simeq 
\frac{1}{m_*}\left( \frac{F_0^S}{1+F_0^S}
-3\frac{F_0^A}{1+F_0^A} 
+4\frac{G_0^S}{1+G_0^S}
-12\frac{G_0^A}{1+G_0^A} 
\right)
\]
becomes negative
close to the MT, finally vanishing as $(m_*)^{-1}$ for $U \to U_{c2}$, while
at weak coupling $t\gg~U\gg~|J|$, 
this quantity is positive ($A_s = U/2 + 2J/3>0$).
This strongly suggests a pairing between quasiparticles.
To check for this instability, we compute the dynamical pair
susceptibility
$
\Pi(\omega) = \langle 0 | \Delta_0^\dagger \left(\omega - H\right)^{-1}
\Delta_0 | 0 \rangle
$, using as pair wave-function the singlet GS of the atomic limit
$\Delta_0 = \sum_a d^\dagger_{0a\uparrow}d^\dagger_{0a\downarrow}$.

\begin{figure}
\centerline{\psfig{bbllx=80pt,bblly=160pt,bburx=510pt,bbury=500pt,%
figure=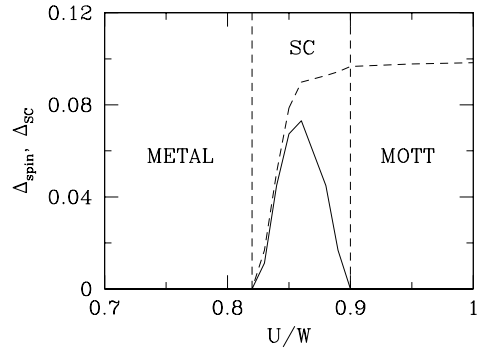,width=60mm,angle=0}}
\caption{Superconducting coupling $\Delta_{SC}$ (solid line) and 
spin gap $\Delta_{spin}$ (dashed line) as computed from the
local spin susceptibility.
The verical dashed lines mark the boundaries of the various phases.
\label{phdsc}
}
\end{figure}

The static limit of the real part of $\Pi(\omega)$, shown in Fig.
\ref{chiekappa} (d), changes sign in a narrow region just before the
MT, turning from negative (stable metal) to positive (unstable). 
This is another clear indication of a
superconducting instability of the FL.
To confirm completely this hypothesis, we finally allow
for a superconducting solution of the DMFT\cite{dmft}.
The lattice model is mapped onto an impurity coupled
to a superconducting bath, or, equivalently, coupled to a normal bath
through a normal and an anomalous term.
\begin{eqnarray}
H_{I}^{S} &=& \sum_{an,\sigma}\epsilon_{an}
c^\dagger_{an\sigma}c^{\phantom{\dagger}}_{an\sigma}
+ \sum_{an,\sigma}
\left(V_{n}d^\dagger_{a\sigma}c^{\phantom{\dagger}}_{an\sigma}+
H.c. \right)
\nonumber\\
& +& \sum_{an}
\left(\Delta_{n}d^\dagger_{a\uparrow}c^{\dagger}_{an\downarrow}+
H.c. \right)
-\mu \sum_{a\sigma} d^\dagger_{a\sigma}d_{a\sigma} + H_{int}.
\label{HAIMSC}
\end{eqnarray}
If the self-consistency equations
\begin{eqnarray*}
&&\sum_{n}\frac{\left|V_{n}\right|^2}{i\omega - \epsilon_{n}}
+\sum_n {\vert \Delta_n\vert^2\over i\omega + \epsilon_n} = t^2 G(i\omega),
\\
&&
-\sum_n {2V_n\epsilon_n\Delta_n\over {\omega^2 + \epsilon_n^2}}=t^2F(i\omega).
\end{eqnarray*} 
stabilize a solution with a finite value of any $\Delta_n$, then a
superconducting solution is found.
We find a stable superconducting GS
in the interval $0.82 < U < 0.9$ (for $J/U=-0.02$)
in close proximity of the MT, as shown in Fig. \ref{phdsc}.
We find that, while the superconducting order parameter
vanishes at the MT, the spin gap of the supercondutor, as computed
from the local spin susceptibility, remains finite,
and joins continuosly into that of the insulator,
as shown in Fig. \ref{phdsc}. Hence neither phases
possesses spin entropy.
A superconducting phase just before the MT is striking, as
there is no sign of, or reason for, such an
instability in the weak-coupling limit, where
the intraorbital $s$-wave scattering amplitude is repulsive.
The anti-adiabatic electron-phonon coupling
implicit in the choice $J<0$, being unretarded, should be
overwhelmed by a large $U\gg |J|$.
As the repulsion $U$ increases, the bare electrons
turn into quasiparticles.
An effective overscreening of $U$ takes then place,
generating the attraction that causes the s-wave pairing. This
is reminiscent of phenomena in other systems, such as
the Hubbard chain with next-nearest neighbor hopping and
the two leg Hubbard ladder\cite{Michele}. Generation of an
attractive interaction in the effective
impurity model, which has an on-site mechanism of singlet formation
competing with the Kondo screening,
is similar to the two impurity Kondo model\cite{Jones&Affleck}.
For reasons of space we must defer pursuit of these analogies to future work.
The heuristic arguments presented earlier suggest that
superconductivity close to a singlet MT should be a phenomenon
of wider generality than the simple model where we have uncovered it.
Experimentally, in the $\langle n\rangle  = 4$
alkali fulleride Rb$_4$C$_{60}$, which can be transformed
by a pressure of about 10 Kbar from its narrow-gap, probably singlet
Mott insulating state,\cite{C60,Fabrizio97} to a metal, no superconductivity
was reported so far down to 0.39 K\cite{kerkoud}. Nonetheless, even
if some features of our model (antiadiabatic 
Jahn-Teller effect, diagonal hopping) are quite
unrealistic, we can expect all the same a narrow 
superconducting pocket sufficiently close to the critical insulator-metal
pressure, at least if the transition is second order, or not too strongly
first order. The small superconducting pocket for $\langle n\rangle = 4$
might be connected to
the much larger one for $\langle n \rangle = 3$, the well-known case of
superconducting fullerides.

We are pleased to acknowledge enlighting discussions with C. Castellani 
and G. Kotliar. M.C. also thanks S. Caprara for helpful suggestions.
We acknowledge support from MURST COFIN99,
and from EU project FULPROP, contract ERBFMRXCT 970155.

\begin{references}
\bibitem{footnotens} As usual, we label by $N_s$ the
total number of ``sites'', so that the number of sites in the
conduction bath is $N_s - 1$.
Notice that, for three-fold degenerate levels, $N_s = 5$
corresponds effectively to $15$ sites.
\end{references}
\end{document}